\documentclass[twocolumn]{aastex62}

\usepackage{amsmath,amssymb,amsfonts}
\usepackage{subfigure}
\usepackage{natbib}
\usepackage{times}
\usepackage{booktabs}
\usepackage{verbatim}
 \newcommand\smone{S_{\text{1}}}
 \newcommand\smtwo{S_{\text{2}}}
  \newcommand\sm{S}
  \newcommand\cmassu{\text{g }\text{cm}^{-2}}
\newcommand\cmass{c_{\text{mass}}}
\newcommand{\rv}[1]{{#1}}
\newcommand{\rvr}[1]{{#1}}

\newcommand\cs{c_{\text{s}}}
\begin{document}
\title{Fast magnetic wave could heat the solar low-beta chromosphere}

\author{Yikang Wang}
\affil{Department of Earth and Planetary Science, The University of Tokyo, 7-3-1 Hongo, Bunkyo, Tokyo 113-0033, Japan}
\author{Takaaki Yokoyama}
\affil{Astronomical Observatory, Kyoto University, Sakyo-ku, Kyoto, 606-8502, Japan}
\author{Haruhisa Iijima}
\affil{Center for Integrated Data Science, Institute for Space-Earth Environmental Research, Nagoya University, Furocho, Chikusa-ku, Nagoya, Aichi 464-8601, Japan}
\affil{National Astronomical Observatory of Japan, 2-21-1 Osawa, Mitaka, Tokyo 181-8588, Japan}
\email{wyk@smail.nju.edu.cn}

\begin{abstract}
Magnetohydrodynamic (MHD) waves are candidates for heating the solar chromosphere, although it is still unclear which mode of the wave is dominant in heating. We perform two-dimensional radiative MHD simulation to investigate the propagation of MHD waves in the quiet region of the solar chromosphere. We identify the mode of the shock waves by using the relationship between gas pressure and magnetic pressure across the shock front and calculate their corresponding heating rate through the entropy jump to obtain a quantitative understanding of the wave heating process in the chromosphere. Our result shows that the fast magnetic wave is significant in heating the low-beta chromosphere. The low-beta fast magnetic waves are generated from high-beta fast acoustic waves via mode conversion crossing the equipartition layer. Efficient mode conversion is achieved by large attacking angles between the propagation direction of the shock waves and the chromospheric magnetic field.
\end{abstract}


\section{Introduction
  \label{sec:introduction:background}}

Mechanical heating is required to maintain the energy balance in the solar chromosphere, as suggested by the temperature difference between the radiative equilibrium atmospheric model \citep{1989ApJ...346.1010A} and the observation-based semi-empirical model \citep{1981ApJS...45..635V}. Waves have been recognized as important contributors to chromospheric heating, although their heating mechanisms are elusive \citep[see][for a review]{2015SSRv..190..103J}. \rvr{While} the propagation of waves in the chromosphere has been well studied from both the observational and theoretical perspectives \rvr{\cite[e.g.,][]{2003ApJ...599..626B,2004A&A...422.1085H,2005ApJ...631.1270H,2008ApJ...680.1542H,2009A&A...508..951V,2011ApJ...743..142H,2012ApJ...755...18V,2013ApJ...764L..11D,2016A&A...585A.110K,2016A&A...590L...3S,2018MNRAS.479.5512K,2020ApJ...890...22A}}, a firm quantitative conclusion is still a distance away. 

\rv{In the chromosphere, physical parameters change drastically, leading to difficulties in studying chromospheric dynamics.} The plasma beta varies in both the vertical and horizontal directions, and waves can change their modes when crossing the equipartition layer \citep{2006RSPTA.364..333C,2019ApJ...881L..21P} where the speed of sound is identical to the Alfv\'en speed. Density stratification also adds to the complexity by increasing the amplitude of the acoustic waves, leading to increased nonlinearity and formation of shocks. 

In the high-beta regions of the chromosphere where the role of the magnetic field could be ignored, the propagation of acoustic waves has been well studied by hydrodynamic simulations with non-local thermodynamic equilibrium (non-LTE) radiative transfer \citep{1995ApJ...440L..29C,1997ApJ...481..500C}. In these studies, waves are generated by longitudinal piston motion. They succeed in reproducing the Ca II spectral profile which agrees with the observations. 

The situation becomes even more complicated in the low-beta chromosphere with the participation of the magnetic field. \cite{2016ApJ...817...94A} and \cite{2016ApJ...829...80B} show that the shock heating rate in the chromosphere is larger than or consistent with the observation-based radiative cooling rate. A similar result is also obtained in \cite{2020ApJ...891..110W} with an improved treatment of the radiative loss term introduced by \cite{2012A&A...539A..39C}. In \cite{2016ApJ...817...94A}, \cite{2016ApJ...829...80B}, and \cite{2020ApJ...891..110W}, waves are generated by artificial transverse torque or transverse motion at the bottom of the flux tube. These studies do not include the effect of waves originating from outside the flux tube. 

\rv{Theoretical studies could be divided into two categories, idealized models and realistic models. \cite{2016ApJ...817...94A}, \cite{2016ApJ...829...80B}, and \cite{2020ApJ...891..110W} are examples of idealized models. The physical process is clear in idealized models, but the results are affected by artificial settings in the model.} On the other hand, there are also realistic models \rvr{\cite[e.g.,][]{2011ApJ...730L..24K,2016A&A...585A...4C,2017ApJ...848...38I,2017Sci...356.1269M}} that aim to include complicated physical processes to approach reality. Realistic models are used to reproduce the synthesized images or spectral profiles for comparison with observations \cite[e.g.,][]{2013ApJ...772...90L,2009ApJ...694L.128L,2019MNRAS.486.4203Q},  but their complexity makes it difficult to understand the underlying elemental physical processes involved in heating. These studies do not focus on the physical processes that occur during wave propagation such as thermalization, nonlinear steepening, or mode conversions in the chromosphere.

\rv{The purpose of our study is to conduct a quantitative investigation on wave heating in the chromosphere. \rv{In particular, previous studies do not focus on the role of the fast magnetic wave in heating the low-beta chromosphere.} We perform a realistic two-dimensional radiative MHD simulation while conducting a detailed investigation on the propagation of waves to estimate the contribution to chromospheric heating by different modes of waves. To achieve this goal, we develop a novel method of automatically identifying the mode of waves and calculating the heating rate due to different modes of waves.}

\section{Numerical model}\label{sec:2}
We use RAMENS code \citep{2016PhDT.........5I,2015ApJ...812L..30I} which solves MHD equations with gravity, heat conduction, equation of state under local thermodynamic equilibrium (LTE) condition, radiative transfer in the photosphere, and approximated radiative loss term in the chromosphere and the corona. The basic equations of the simulation are the same as those in \cite{2017ApJ...848...38I}. One could refer to \cite{2016PhDT.........5I} for a detailed description of this code. We modified the original RAMENS code by replacing the treatment of the chromospheric radiative loss term with the improved recipe developed by  \cite{2012A&A...539A..39C}.

The simulation domain is a 16 Mm $\times$ 16 Mm two-dimensional square extending from $-$ 2 Mm below the photosphere to 14 Mm in the corona with a uniform grid spacing of 8.5 km. The temperature of the corona is 1 MK which is maintained by the top boundary condition. The initial magnetic field is vertical and has a strength of 6 G. We start with a plane parallel atmosphere in the hydrostatic equilibrium state, though this setup does not strongly influence the later results obtained after the well-developed magneto-convections. The data analyzed cover 1000 s of the simulation which is approximately 10 times the transit time for acoustic waves in the chromosphere.

\section{Shock identification and heating rate calculation}\label{sec:21}
Our study focuses on wave heating in the low-beta chromosphere. Comparing this mechanism with other possible heating mechanisms (e.g., reconnection and turbulence with ambipolar diffusion), there is observational evidence showing that waves can carry enough energy for chromospheric heating \citep{2010ApJ...723L.134B}. Waves are generated by photospheric convection and steepen to shocks as they propagate upward in the chromosphere. To estimate the shock-heating rate, we identify the shock front in the chromosphere, determine the mode of each shock, and calculate the corresponding heating rate.
The positions of the shock fronts are identified by the local minimum of $\nabla \cdot \mathbf{V}$ with
\begin{equation}\label{eq:sel22}
-\nabla \cdot \mathbf{V} \ge  c_{\text{th}}  (\cs/\Delta x),
\end{equation}
where $c_{\text{th}}$ is a parameter indicating the threshold for identification, $\cs$ is the speed of sound, and $\Delta x$ is the grid size. The value of the parameter $c_{\text{th}}$ should depend on the shock-capturing quality of the numerical scheme and was taken to be $c_{\text{th}}=0.25$ in this study \citep[see Appendix in][]{2020ApJ...891..110W}. 

\begin{figure}
    \centering
    \includegraphics[width=7cm]{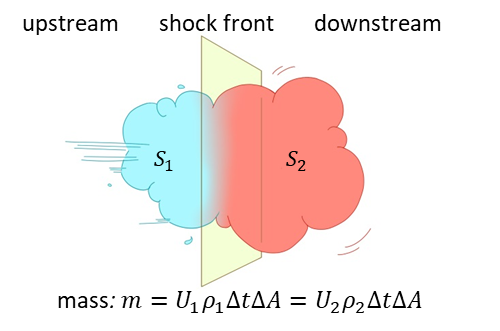}
    \caption{\rvr{A schematic figure showing the calculation of thermal energy flux. $t$ is time. $A$ is the area on the shock front. $m=U_1 \rho_1\Delta t\Delta A=U_2 \rho_2\Delta t\Delta A$ is the mass of plasma that crosses the shock front. Color in the upstream and the downstream regions denotes the value of entropy per unit mass (red: higher value, blue: lower value).  $\Delta Q_{\text{m}}=T\Delta\sm$ is the increment of thermal energy per unit mass. Thus, we can obtain the thermal flux by $\Delta F_{\text{th}}=m\Delta Q_{\text{m}}/(\Delta t\Delta A)$}.}
    \label{fig:sc}
\end{figure}

\rv{The heating rate at the shock front is calculated using the following steps}. First, we extract the density, temperature, velocity, gas pressure, and magnetic pressure along the direction of propagation which is assumed to be identical to the direction of the gradient of the total pressure. The upstream and downstream quantities of the detected shock are determined as the first local maximum and minimum of $\partial^2 V_l/\partial l^2$ beside $l = l_{\text{c}}$ where $V_l$ is the velocity along the direction of propagation, $l$ is the distance along the direction of propagation, and $l_{\text{c}}$ is the position of the shock front. The upstream side is determined by the side with the lower density. We estimate the increment of the thermal energy flux at the shock front:
\begin{equation}\label{eq:ql}
\Delta F_{\text{th}} = U_1 \rho_1 T_1 (\smone-\smtwo),
\end{equation}
where $\Delta F_{\text{th}}$ is the increase in the thermal energy flux. Subscripts 1 and 2 denote the physical parameters that are sampled at the upstream and downstream region, respectively. $U$ is the shock-normal velocity in the shock rest frame, $T$ is the temperature, and $\sm$ is the entropy per unit mass. $U_1$ is calculated by mass conservation, $U_1 \rho_1 = U_2 \rho_2$, and the velocity relationship in different frames of reference, $v_1 - v_2 = U_1 - U_2$ where $\rho$ is the density and $v$ is the shock-normal velocity in the laboratory frame \rvr{(see Figure \ref{fig:sc} for a schematic plot)}. To estimate the heating rate per unit volume, we assume that the heating is evenly distributed in the volume of one grid point at the shock front. As a result, the heating rate per unit volume is calculated by
\begin{equation}\label{eq:cal}
Q_{\text{heat}}=\Delta F_{\text{th}}/w_{\text{shock}},
\end{equation}
where $w_{\text{shock}}$ is the width of the shock wave. Although the actual thickness in the real shocks should be given by the microscopic dissipation process, we here use the grid spacing $w_{\text{shock}} = \Delta x$ for convenience. \rvr{The heating rate is calculated each time step. We assume that the heating rate at a fixed position do not change within one time step}. Note that the spatially integrated amount of $Q_{\text{heat}}$ is independent of the choice of $w_{\text{shock}}$ and is used only for the later discussion.

\rv{Finally, we determine the mode of each shock wave \rvr{by checking whether the gas pressure and the magnetic pressure across the shock front change in the same direction}. The sign of $\int (\partial P_g / \partial l)(\partial P_m/ \partial l) \text{d} l$ across the shock front is used to determine whether it is a fast shock (positive value) or slow shock (negative value), where $P_g$ is the gas pressure and $P_m$ is the magnetic pressure. \rvr{We do not use phase speed to determine the mode of waves since it is difficult to obtain the local fast speed and slow speed in the dynamic chromosphere.}} 

\section{Results} \label{sec:3}
\begin{figure*}
    \centering
    \includegraphics[width=15cm]{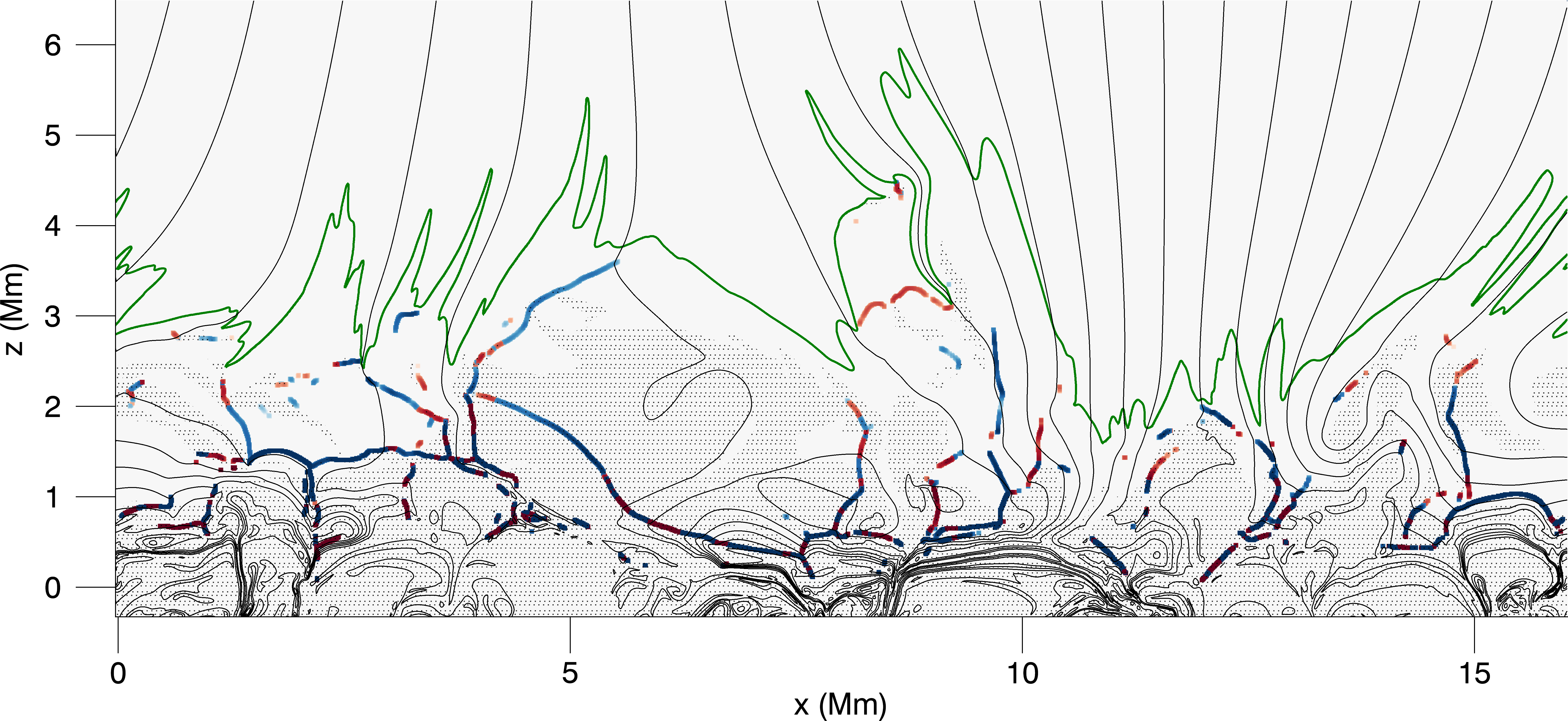}
    \caption{Snapshot of the simulation result with shock identification. The green line marks the position of the transition region (characterized by $T = 10^4$ K). The black solid lines are magnetic field lines. The gray shadow indicates the region where speed of sound is larger than the Alfv\'en speed. Identified shocks are plotted in blue (fast shock) and red (slow shock). Only a part of the simulation domain is shown in this figure. 
    (An animation of this figure is available.)}
    \label{fig:fig1}
\end{figure*}

Figure \ref{fig:fig1} shows the identified shock fronts in the dynamic simulation of solar chromosphere. Waves are generated by photospheric convection and they steepen to shocks in the chromosphere. Shocks dissipate their energy continuously in the chromosphere. A number of shocks gradually become undetectable during their propagation due to dissipation. When shocks impinge on the transition region, they drive the upward motion of the transition region that forms spicules.

\begin{figure}
    \centering
    \includegraphics[width=5cm]{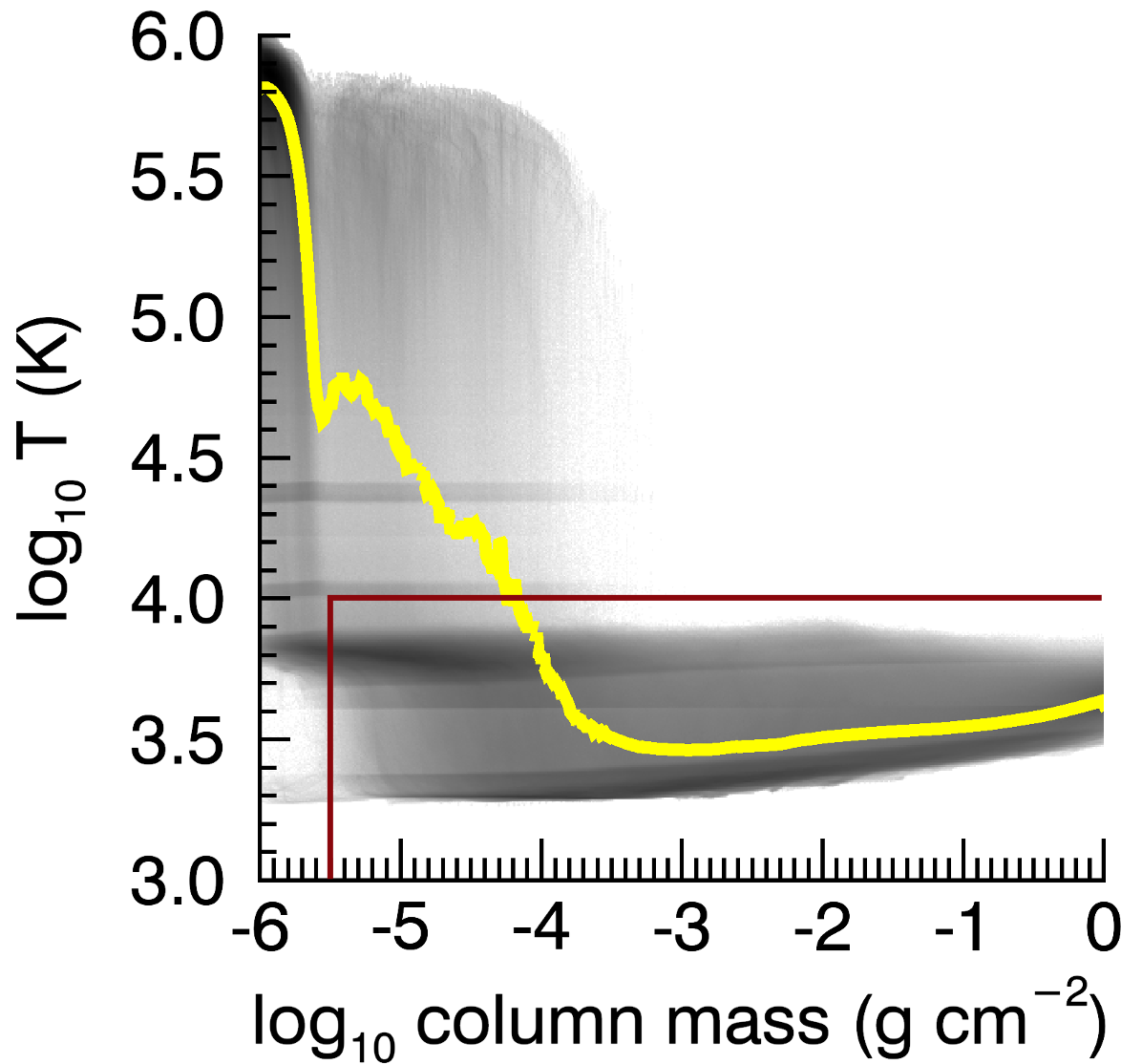}
    \caption{Joint probability density distribution of the temperature and the column mass. The yellow line shows the average temperature at each column mass. The brown lines show the threshold values for chromospheric plasma at $\cmass=10^{-5.5}$ $\cmassu$ and $T = 10^4$ K.}
    \label{fig:fig2-1}
\end{figure}

\rv{We focus on the low-beta chromospheric plasma. Due to the large deformation of the transition region by the spicules, we cannot distinguish the chromosphere and the corona using a simple threshold on the geometrical height. The \rvr{low-beta} chromospheric plasma is defined by the following criteria}: (1) $\cmass> 10^{-5.5}$ $\cmassu$, (2) $T< 10^4$ K, and (3) Alfv\'en speed is larger than sound speed. The variable $\cmass$ is the column mass, and $\cmass(z)=\int_z^{z_{\text{top}}} \rho(z') \text{d} z'$, where $z_{\text{top}}$ is the height of the top of the simulation box. The temperature and column mass threshold are used to exclude coronal plasma. The values of the thresholds are chosen from the joint probability density distribution of the temperature and the column mass (Figure \ref{fig:fig2-1}).  

\begin{figure}
    \centering
    \includegraphics[width=8cm]{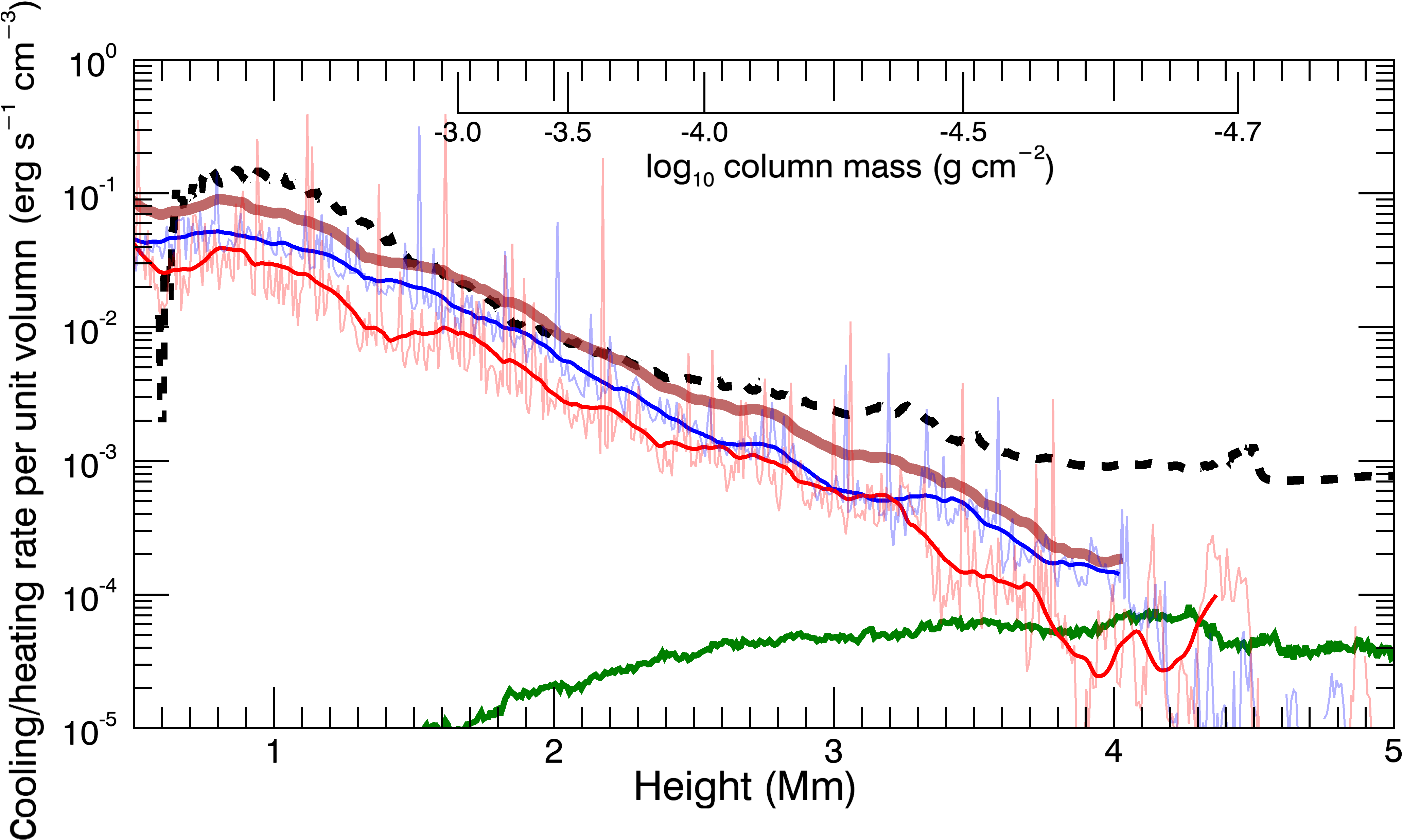}
    \caption{Heating and radiative loss rate of the low-beta chromospheric plasma as a function of height. The black dashed line is the radiative loss rate in the simulation. The brown line is the sum of the heating rates due to fast and slow shocks. The blue solid line is the fast wave heating rate. The red solid line is the slow wave heating rate. For the blue and red lines, the thin lines with perturbation are the results that are directly calculated from the simulation; we also smooth the results with a Savitsky--Golay filter and plot them in thick lines. The green line represents the heating rate due to heat conduction. The average column mass at each height is shown in the secondary axis. \rvr{Only heating and cooling in the low-beta regions are included in this figure.}}
    \label{fig:fig2}
\end{figure}

\rv{The time and horizontal averaged radiative loss rate and heating rate of the low-beta chromospheric plasma are shown in Figure \ref{fig:fig2}}. It is shown that the shock heating is well balanced with radiative cooling below 2.5 Mm. At locations higher than 2.5 Mm, the energy balance is gradually disrupted due to the formation of spicules (in the presence of spicules, the energy balance at a fixed position is determined by the entropy flow carried by them). 

\rv{Where do these fast mode waves in the low-beta regions originate? }We find that low-beta fast magnetic waves originate from high-beta fast acoustic waves through mode conversion. An example of mode conversion is shown in Figure \ref{fig:fig3}. Mode conversion occurs when fast acoustic waves propagate from the high-beta region to the low-beta region and cross the equipartition layer. An attacking angle (the angle between the wavevector and the magnetic field) close to $90^{\circ}$ will result in a larger conversion rate \citep{2006RSPTA.364..333C,2019ApJ...881L..21P}.

  \begin{figure*}
    \centering
    \includegraphics[width=13cm]{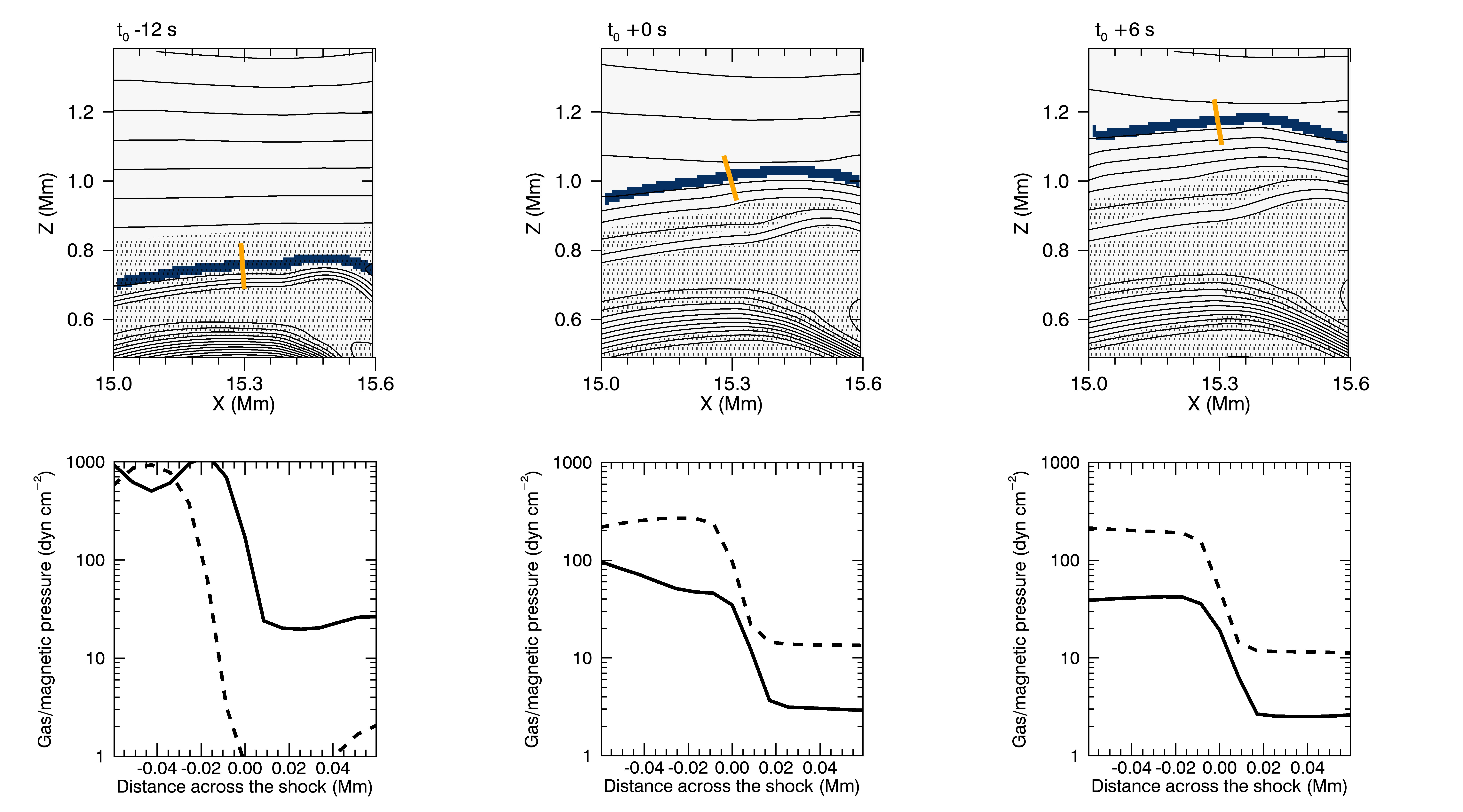}
    \caption{Example of a fast acoustic wave to fast magnetic wave mode conversion. The upper panels show the time evolution (from left to right: $t_0 - 12$ s, $t_0$, and $t_0 + 6$ s, where $t_0$ is the time of the snapshot shown in Figure \ref{fig:fig1}). In the upper panels, gray lines represent magnetic field lines. The blue line shows the position of a fast shock. Shadows mark the region where the speed of sound is larger than the Alfv\'en speed. Orange lines mark the position of slices used in the lower panels. The lower panels show the distribution of the gas pressure (solid line) and magnetic pressure (dashed line) across the shock front. In each panel, the horizontal axis is the distance along the slice, in which zero corresponds to the location of the shock front.}
    \label{fig:fig3}
\end{figure*}

    \label{fig:fig3}

\section{Discussion
  \label{sec:4}}

The propagation of waves in MHD simulation with an idealized setting is also \rvr{carried on in previous researches \citep{2005ApJ...631.1270H,2008ApJ...680.1542H,2009A&A...508..951V,2012ApJ...755...18V}. These studies mainly focus on waves that originate inside a flux tube. For these waves, as they propagate upwards, they propagate along the magnetic field lines thus the attacking angle is small and mode conversion is less efficient}. \cite{2008ApJ...680.1542H} do mention the waves that originate outside a flux tube could generate fast magnetic waves in the flux tube through mode conversion but they do not discuss the heating by the fast magnetic waves in detail. Our result shows that, with quantification of the heating rate, fast waves do play a role in heating the low-beta chromosphere. 


\rvr{\cite{2006ApJ...653..739K} show that refraction could affect the propagation of fast waves and prevent their efficient energy transport to the chromosphere. They focus on waves inside a strong flux tube (sunspots). On the other hand, in our simulation, fast waves in the regions between two flux tubes are less affected by refraction since there is no substantial horizontal gradient of fast speed in these regions. In addition, the intensity of magnetic field in the flux tube is weaker in our simulation which also reduces the horizontal gradient of fast speed.}

Our simulation shows that shock heating is the dominant heating process in the chromosphere. This result is consistent with those from previous studies. However, the wave modes contributing to heating are different. In \cite{2016ApJ...817...94A}, \cite{2020ApJ...891..110W} and \rvr{\cite{2004A&A...422.1085H}}, transverse waves at the foot of a low-beta flux tube undergo nonlinear mode coupling and generate slow acoustic waves. They steepen to shocks which dissipate and contribute to chromospheric heating. In our simulation, Alfv\`en waves vanish because of the two-dimensional geometry. As a result, the nonlinear mode coupling is also absent.

As fast waves propagate like an expanding sphere, the strongest perturbation of the vertical velocity appears at the top of the sphere, whereas compression of the vertical magnetic field appears at the lateral sides. In our simulation, the background magnetic field is 6 G, mimicking the quiet sun region. The resultant intensity of the magnetic field perturbation in the chromosphere could be as large as 10--20 G. The combination of vertical velocity and vertical magnetic field perturbation can be used as a signal of the fast wave. Such a signal can hopefully be detected by next-generation solar telescopes such as Daniel K. Inouye Solar Telescope \citep[DKIST;][]{2020SoPh..295..172R} and Chinese Giant Solar Telescope \citep[CGST;][]{2011ASInC...2...31D}.

\rvr{In order to investigate effect of the topology of magnetic field line, we carry on another simulation with the same initial and boundary condition described in Section \ref{sec:2}. The only difference is that we increase the intensity of the initial background magnetic field from 6 G to 20 G. In this new setting, the magnetic field lines are less inclined which results in smaller attacking angle for waves that propagate upward. We find that the percentage of heating by slow wave increases, especially in the higher part of the chromosphere characterized by $\cmass<10^{-4.2}$ $\cmassu$. However, our main result that fast magnetic shock waves play a significant role in heating the low-beta chromosphere remains unchanged.}

\rvr{Our study is limited in the quiet region. In sunspots, observations show that wave energy is insufficient for chromospheric heating \citep{2011ApJ...735...65F}. In these regions, other effects related with magnetic field such as reconnection should be taken into consideration.}

In this study, the ambipolar diffusion and dynamic ionization of hydrogen are not considered. The dissipation of ambipolar diffusion could lead to substantial heating locally \rvr{\citep{2012ApJ...747...87K,2016ApJ...819L..11S,2017Sci...356.1269M,2019ApJ...871....3S}.} On the other hand, \cite{2016ApJ...817...94A} compare the time-averaged heating rate resulting from ambipolar diffusion and shock dissipation and find that shock heating is much stronger than the heating resulting from ambipolar diffusion. \cite{2007A&A...473..625L} compare simulations with the LTE assumption and dynamic ionization. It is shown that in the simulation with dynamic ionization, shock temperatures are higher and the intershock temperatures are lower than in the simulation with the LTE assumption. This effect could affect the measurement of the entropy jump. Moreover, dynamic ionization is important to determine the electron and the ion number density and will further affect the estimation of ambipolar diffusion, especially when the ionization degree is low. Further studies that compare shock heating, turbulence heating \citep[][]{2011ApJ...736....3V}, and ambipolar diffusion \rvr{\citep[][]{2005A&A...442.1091L,2018A&A...618A..87K,2020ApJ...889...95M,2020A&A...642A.220G}} in realistic simulations are expected to be conducted in the future.

\section{Conclusion
  \label{chap:summary}}
We perform a two-dimensional MHD simulation to study the propagation of MHD waves in the chromosphere. We identify the mode of the shock waves in the chromosphere, calculate the heating rate from the entropy jump, and find that the heating rate balances with the radiative loss. Fast magnetic shock waves play a significant role in heating the low-beta chromosphere. These low-beta fast magnetic waves are generated by mode conversion.

\rvr{We acknowledge the referee for valuable comments.} The authors thank M. Carlsson for providing numerical tables for the recipe of the chromospheric radiative loss. \rvr{The authors thank B. Yu's assistance in making Figure \ref{fig:sc}.} Numerical computations were carried out on Cray XC50 at Center for Computational Astrophysics, National Astronomical Observatory of Japan. T.Y. is supported by JSPS KAKENHI grant No. 15H03640, No. 20KK0072, and No. 21H01124. H.I. is supported by JSPS KAKENHI grant No. 19K14756.
\clearpage
\bibliographystyle{apj}
\bibliography{reference}

\begin{thebibliography}{46}
\expandafter\ifx\csname natexlab\endcsname\relax\def\natexlab#1{#1}\fi

\bibitem[{{Abbasvand} {et~al.}(2020){Abbasvand}, {Sobotka}, {Heinzel},
  {{\v{S}}vanda}, {Jur{\v{c}}{\'a}k}, {del Moro}, \&
  {Berrilli}}]{2020ApJ...890...22A}
{Abbasvand}, V., {Sobotka}, M., {Heinzel}, P., {{\v{S}}vanda}, M.,
  {Jur{\v{c}}{\'a}k}, J., {del Moro}, D., \& {Berrilli}, F. 2020, \apj, 890, 22

\bibitem[{{Anderson} \& {Athay}(1989)}]{1989ApJ...346.1010A}
{Anderson}, L.~S., \& {Athay}, R.~G. 1989, \apj, 346, 1010

\bibitem[{{Arber} {et~al.}(2016){Arber}, {Brady}, \&
  {Shelyag}}]{2016ApJ...817...94A}
{Arber}, T.~D., {Brady}, C.~S., \& {Shelyag}, S. 2016, ApJ, 817, 94

\bibitem[{{Bello Gonz{\'a}lez} {et~al.}(2010){Bello Gonz{\'a}lez}, {Franz},
  {Mart{\'{\i}}nez Pillet}, {Bonet}, {Solanki}, {del Toro Iniesta}, {Schmidt},
  {Gandorfer}, {Domingo}, {Barthol}, {Berkefeld}, \&
  {Kn{\"o}lker}}]{2010ApJ...723L.134B}
{Bello Gonz{\'a}lez}, N., {et~al.} 2010, ApJL, 723, L134

\bibitem[{{Bogdan} {et~al.}(2003){Bogdan}, {Carlsson}, {Hansteen}, {McMurry},
  {Rosenthal}, {Johnson}, {Petty-Powell}, {Zita}, {Stein}, {McIntosh}, \&
  {Nordlund}}]{2003ApJ...599..626B}
{Bogdan}, T.~J., {et~al.} 2003, \apj, 599, 626

\bibitem[{{Brady} \& {Arber}(2016)}]{2016ApJ...829...80B}
{Brady}, C.~S., \& {Arber}, T.~D. 2016, ApJ, 829, 80

\bibitem[{{Cally}(2006)}]{2006RSPTA.364..333C}
{Cally}, P.~S. 2006, Philosophical Transactions of the Royal Society of London
  Series A, 364, 333

\bibitem[{{Carlsson} {et~al.}(2016){Carlsson}, {Hansteen}, {Gudiksen},
  {Leenaarts}, \& {De Pontieu}}]{2016A&A...585A...4C}
{Carlsson}, M., {Hansteen}, V.~H., {Gudiksen}, B.~V., {Leenaarts}, J., \& {De
  Pontieu}, B. 2016, \aap, 585, A4

\bibitem[{{Carlsson} \& {Leenaarts}(2012)}]{2012A&A...539A..39C}
{Carlsson}, M., \& {Leenaarts}, J. 2012, A\&A, 539, A39

\bibitem[{{Carlsson} \& {Stein}(1995)}]{1995ApJ...440L..29C}
{Carlsson}, M., \& {Stein}, R.~F. 1995, \apjl, 440, L29

\bibitem[{{Carlsson} \& {Stein}(1997)}]{1997ApJ...481..500C}
---. 1997, \apj, 481, 500

\bibitem[{{de la Cruz Rodr{\'\i}guez} {et~al.}(2013){de la Cruz
  Rodr{\'\i}guez}, {De Pontieu}, {Carlsson}, \& {Rouppe van der
  Voort}}]{2013ApJ...764L..11D}
{de la Cruz Rodr{\'\i}guez}, J., {De Pontieu}, B., {Carlsson}, M., \& {Rouppe
  van der Voort}, L.~H.~M. 2013, \apjl, 764, L11

\bibitem[{{Deng} \& {CGST Group}(2011)}]{2011ASInC...2...31D}
{Deng}, Y.~Y., \& {CGST Group}. 2011, in Astronomical Society of India
  Conference Series, Vol.~2, Astronomical Society of India Conference Series,
  31--36

\bibitem[{{Felipe} {et~al.}(2011){Felipe}, {Khomenko}, \&
  {Collados}}]{2011ApJ...735...65F}
{Felipe}, T., {Khomenko}, E., \& {Collados}, M. 2011, \apj, 735, 65

\bibitem[{{Gonz{\'a}lez-Morales} {et~al.}(2020){Gonz{\'a}lez-Morales},
  {Khomenko}, {Vitas}, \& {Collados}}]{2020A&A...642A.220G}
{Gonz{\'a}lez-Morales}, P.~A., {Khomenko}, E., {Vitas}, N., \& {Collados}, M.
  2020, \aap, 642, A220

\bibitem[{{Hasan} \& {Ulmschneider}(2004)}]{2004A&A...422.1085H}
{Hasan}, S.~S., \& {Ulmschneider}, P. 2004, \aap, 422, 1085

\bibitem[{{Hasan} \& {van Ballegooijen}(2008)}]{2008ApJ...680.1542H}
{Hasan}, S.~S., \& {van Ballegooijen}, A.~A. 2008, \apj, 680, 1542

\bibitem[{{Hasan} {et~al.}(2005){Hasan}, {van Ballegooijen}, {Kalkofen}, \&
  {Steiner}}]{2005ApJ...631.1270H}
{Hasan}, S.~S., {van Ballegooijen}, A.~A., {Kalkofen}, W., \& {Steiner}, O.
  2005, \apj, 631, 1270

\bibitem[{{Heggland} {et~al.}(2011){Heggland}, {Hansteen}, {De Pontieu}, \&
  {Carlsson}}]{2011ApJ...743..142H}
{Heggland}, L., {Hansteen}, V.~H., {De Pontieu}, B., \& {Carlsson}, M. 2011,
  \apj, 743, 142

\bibitem[{{Iijima}(2016)}]{2016PhDT.........5I}
{Iijima}, H. 2016, PhD thesis, Department of Earth and Planetary Science,
  School of Science, The University of Tokyo, Japan

\bibitem[{{Iijima} \& {Yokoyama}(2015)}]{2015ApJ...812L..30I}
{Iijima}, H., \& {Yokoyama}, T. 2015, \apjl, 812, L30

\bibitem[{{Iijima} \& {Yokoyama}(2017)}]{2017ApJ...848...38I}
---. 2017, \apj, 848, 38

\bibitem[{{Jess} {et~al.}(2015){Jess}, {Morton}, {Verth}, {Fedun}, {Grant}, \&
  {Giagkiozis}}]{2015SSRv..190..103J}
{Jess}, D.~B., {Morton}, R.~J., {Verth}, G., {Fedun}, V., {Grant}, S.~D.~T., \&
  {Giagkiozis}, I. 2015, SSRev, 190, 103

\bibitem[{{Kato} {et~al.}(2011){Kato}, {Steiner}, {Steffen}, \&
  {Suematsu}}]{2011ApJ...730L..24K}
{Kato}, Y., {Steiner}, O., {Steffen}, M., \& {Suematsu}, Y. 2011, \apjl, 730,
  L24

\bibitem[{{Kayshap} {et~al.}(2018){Kayshap}, {Murawski}, {Srivastava},
  {Musielak}, \& {Dwivedi}}]{2018MNRAS.479.5512K}
{Kayshap}, P., {Murawski}, K., {Srivastava}, A.~K., {Musielak}, Z.~E., \&
  {Dwivedi}, B.~N. 2018, \mnras, 479, 5512

\bibitem[{{Khomenko} \& {Collados}(2006)}]{2006ApJ...653..739K}
{Khomenko}, E., \& {Collados}, M. 2006, \apj, 653, 739

\bibitem[{{Khomenko} \& {Collados}(2012)}]{2012ApJ...747...87K}
---. 2012, \apj, 747, 87

\bibitem[{{Khomenko} {et~al.}(2018){Khomenko}, {Vitas}, {Collados}, \& {de
  Vicente}}]{2018A&A...618A..87K}
{Khomenko}, E., {Vitas}, N., {Collados}, M., \& {de Vicente}, A. 2018, \aap,
  618, A87

\bibitem[{{Kontogiannis} {et~al.}(2016){Kontogiannis}, {Tsiropoula}, \&
  {Tziotziou}}]{2016A&A...585A.110K}
{Kontogiannis}, I., {Tsiropoula}, G., \& {Tziotziou}, K. 2016, \aap, 585, A110

\bibitem[{{Leake} {et~al.}(2005){Leake}, {Arber}, \&
  {Khodachenko}}]{2005A&A...442.1091L}
{Leake}, J.~E., {Arber}, T.~D., \& {Khodachenko}, M.~L. 2005, \aap, 442, 1091

\bibitem[{{Leenaarts} {et~al.}(2009){Leenaarts}, {Carlsson}, {Hansteen}, \&
  {Rouppe van der Voort}}]{2009ApJ...694L.128L}
{Leenaarts}, J., {Carlsson}, M., {Hansteen}, V., \& {Rouppe van der Voort}, L.
  2009, \apjl, 694, L128

\bibitem[{{Leenaarts} {et~al.}(2007){Leenaarts}, {Carlsson}, {Hansteen}, \&
  {Rutten}}]{2007A&A...473..625L}
{Leenaarts}, J., {Carlsson}, M., {Hansteen}, V., \& {Rutten}, R.~J. 2007, \aap,
  473, 625

\bibitem[{{Leenaarts} {et~al.}(2013){Leenaarts}, {Pereira}, {Carlsson},
  {Uitenbroek}, \& {De Pontieu}}]{2013ApJ...772...90L}
{Leenaarts}, J., {Pereira}, T.~M.~D., {Carlsson}, M., {Uitenbroek}, H., \& {De
  Pontieu}, B. 2013, \apj, 772, 90

\bibitem[{{Mart{\'\i}nez-Sykora} {et~al.}(2017){Mart{\'\i}nez-Sykora}, {De
  Pontieu}, {Hansteen}, {Rouppe van der Voort}, {Carlsson}, \&
  {Pereira}}]{2017Sci...356.1269M}
{Mart{\'\i}nez-Sykora}, J., {De Pontieu}, B., {Hansteen}, V.~H., {Rouppe van
  der Voort}, L., {Carlsson}, M., \& {Pereira}, T.~M.~D. 2017, Science, 356,
  1269

\bibitem[{{Mart{\'\i}nez-Sykora} {et~al.}(2020){Mart{\'\i}nez-Sykora},
  {Leenaarts}, {De Pontieu}, {N{\'o}brega-Siverio}, {Hansteen}, {Carlsson}, \&
  {Szydlarski}}]{2020ApJ...889...95M}
{Mart{\'\i}nez-Sykora}, J., {Leenaarts}, J., {De Pontieu}, B.,
  {N{\'o}brega-Siverio}, D., {Hansteen}, V.~H., {Carlsson}, M., \&
  {Szydlarski}, M. 2020, \apj, 889, 95

\bibitem[{{Pennicott} \& {Cally}(2019)}]{2019ApJ...881L..21P}
{Pennicott}, J.~D., \& {Cally}, P.~S. 2019, \apjl, 881, L21

\bibitem[{{Quintero Noda} {et~al.}(2019){Quintero Noda}, {Iijima}, {Katsukawa},
  {Shimizu}, {Carlsson}, {de la Cruz Rodr{\'\i}guez}, {Ruiz Cobo}, {Orozco
  Su{\'a}rez}, {Oba}, {Anan}, {Kubo}, {Kawabata}, {Ichimoto}, \&
  {Suematsu}}]{2019MNRAS.486.4203Q}
{Quintero Noda}, C., {et~al.} 2019, \mnras, 486, 4203

\bibitem[{{Rimmele} {et~al.}(2020){Rimmele}, {Warner}, {Keil}, {Goode},
  {Kn{\"o}lker}, {Kuhn}, {Rosner}, {McMullin}, {Casini}, {Lin}, {W{\"o}ger},
  {von der L{\"u}he}, {Tritschler}, {Davey}, {de Wijn}, {Elmore}, {Fehlmann},
  {Harrington}, {Jaeggli}, {Rast}, {Schad}, {Schmidt}, {Mathioudakis},
  {Mickey}, {Anan}, {Beck}, {Marshall}, {Jeffers}, {Oschmann}, {Beard},
  {Berst}, {Cowan}, {Craig}, {Cross}, {Cummings}, {Donnelly}, {de Vanssay},
  {Eigenbrot}, {Ferayorni}, {Foster}, {Galapon}, {Gedrites}, {Gonzales},
  {Goodrich}, {Gregory}, {Guzman}, {Guzzo}, {Hegwer}, {Hubbard}, {Hubbard},
  {Johansson}, {Johnson}, {Liang}, {Liang}, {McQuillen}, {Mayer}, {Newman},
  {Onodera}, {Phelps}, {Puentes}, {Richards}, {Rimmele}, {Sekulic}, {Shimko},
  {Simison}, {Smith}, {Starman}, {Sueoka}, {Summers}, {Szabo}, {Szabo},
  {Wampler}, {Williams}, \& {White}}]{2020SoPh..295..172R}
{Rimmele}, T.~R., {et~al.} 2020, \solphys, 295, 172

\bibitem[{{Santamaria} {et~al.}(2016){Santamaria}, {Khomenko}, {Collados}, \&
  {de Vicente}}]{2016A&A...590L...3S}
{Santamaria}, I.~C., {Khomenko}, E., {Collados}, M., \& {de Vicente}, A. 2016,
  \aap, 590, L3

\bibitem[{{Shelyag} {et~al.}(2016){Shelyag}, {Khomenko}, {de Vicente}, \&
  {Przybylski}}]{2016ApJ...819L..11S}
{Shelyag}, S., {Khomenko}, E., {de Vicente}, A., \& {Przybylski}, D. 2016,
  \apjl, 819, L11

\bibitem[{{Soler} {et~al.}(2019){Soler}, {Terradas}, {Oliver}, \&
  {Ballester}}]{2019ApJ...871....3S}
{Soler}, R., {Terradas}, J., {Oliver}, R., \& {Ballester}, J.~L. 2019, \apj,
  871, 3

\bibitem[{{van Ballegooijen} {et~al.}(2011){van Ballegooijen}, {Asgari-Targhi},
  {Cranmer}, \& {DeLuca}}]{2011ApJ...736....3V}
{van Ballegooijen}, A.~A., {Asgari-Targhi}, M., {Cranmer}, S.~R., \& {DeLuca},
  E.~E. 2011, \apj, 736, 3

\bibitem[{{Vernazza} {et~al.}(1981){Vernazza}, {Avrett}, \&
  {Loeser}}]{1981ApJS...45..635V}
{Vernazza}, J.~E., {Avrett}, E.~H., \& {Loeser}, R. 1981, ApJS, 45, 635

\bibitem[{{Vigeesh} {et~al.}(2012){Vigeesh}, {Fedun}, {Hasan}, \&
  {Erd{\'e}lyi}}]{2012ApJ...755...18V}
{Vigeesh}, G., {Fedun}, V., {Hasan}, S.~S., \& {Erd{\'e}lyi}, R. 2012, \apj,
  755, 18

\bibitem[{{Vigeesh} {et~al.}(2009){Vigeesh}, {Hasan}, \&
  {Steiner}}]{2009A&A...508..951V}
{Vigeesh}, G., {Hasan}, S.~S., \& {Steiner}, O. 2009, \aap, 508, 951

\bibitem[{{Wang} \& {Yokoyama}(2020)}]{2020ApJ...891..110W}
{Wang}, Y., \& {Yokoyama}, T. 2020, \apj, 891, 110

\end{thebibliography}




\end{document}